\documentclass[11pt]{article}
\begin{document}
\title{
Discrete zero curvature representations and infinitely many conservation laws
for several 2+1 dimensional lattice hierarchies}
\author{
Zuo-nong Zhu \\
Department of Mathematics, Shanghai Jiao Tong University,\\
Shanghai, 200030, P.R. China, E-mail: znzhu@online.sh.cn}
\date{}
\maketitle
\begin{abstract}
In this article, several 2+1 dimensional lattice hierarchies proposed by Blaszak and Szum [J. Math. Phys. {\bf 42}, 225(2001)] are further investigated. We first describe
their discrete zero curvature representations. Then, by means of solving the corresponding discrete spectral equation, we demonstrate the existence of infinitely many conservation laws for them and obtain the corresponding conserved densities and associated fluxes formulaically. Thus, their integrability is further confirmed.
\end{abstract}
\setcounter{section}{0}
\setcounter{equation}{0}
\section{Introduction}
In recent years there has been wide interest in the study of nonlinear integrable lattice systems.  It is well known that discrete lattice systems not only possess 
rich mathematical structures, but also have many applications in science, such as mathematical physics, numerical analysis, statistical physics, quantum physics, etc. Much research has been
obtained on $1+1$ dimensional lattice systems. The most famous and well studied 1+1 dimensional integrable lattice systems are the Toda lattice and the Volterra lattice. It has been shown that they possess all integrable properties, such as the Lax pairs, the Hamiltonian structures, infinitely many conservation laws, the B\"{a}cklund transformation, and soliton solutions [1-8]. The $2+1$ dimensional integrable lattice systems have also
been studied such as $2+1$ dimensional Toda lattice
\begin{eqnarray}
\frac{\partial^2ln v(n)}{\partial t^2}-\frac{\partial^2ln v(n)}{\partial t\partial y}
=v(n+1)-2v(n)+v(n-1)
\end{eqnarray}
Equation (1.1) possesses the Lax representation, the B\"{a}cklund transformation and soliton solutions. It admits the infinitely many conservation laws and the Hamiltonian structure [9, 10]. However, comparing with the $1+1$ dimensional lattice systems, there is less work on $2+1$ dimensional integrable lattice systems. Recently, by means of the so-called central extension procedure and operand formalism,
Blaszak and Szum constructed some new 2+1 dimensional integrable hierarchies which have the Hamiltonian structures.
Let's recall some results presented by Blaszak and Szum in Ref. [11].
Considering finite field Lax operator $\tilde{L}$ in the form
\begin{eqnarray}
\tilde{L}=\varepsilon^{m+\alpha}+u_{m+\alpha-1}\varepsilon^{m+\alpha-1}+...+u_{\alpha}\varepsilon^{\alpha}, \qquad 0\geq \alpha >-m.
\end{eqnarray}
where operator $\varepsilon$ is defined by
\begin{eqnarray*}
\varepsilon^iu(n)=(E^iu(n))\varepsilon^i=u(n+i)\varepsilon^i, \qquad i\in Z,
\end{eqnarray*}
$E$ acts on u(n) as a unit shift: Eu(n)=u(n+1), $n\in Z$, and setting
\begin{eqnarray}
\bigtriangledown C_i=\sum_{i\geq j}a_j(n)\varepsilon^j, \qquad i=1,2,...,
\end{eqnarray}
where the function parameters $a_j$ are determined from the following equation succesively via the recurrent procedure:
\begin{eqnarray}
[\bigtriangledown C_i, \tilde{L}-\partial_y]=0,
\end{eqnarray}
then, the following 2+1 dimensional lattice hierarchy is proposed,
\begin{eqnarray}
\tilde{L}_{t_i}=[P_{\geq 0}\bigtriangledown C_i, \tilde{L}-\partial_y],
\end{eqnarray}
where 
\begin{eqnarray}
P_{\geq 0}\bigtriangledown C_i=\sum_{j\geq 0}a_j(n)\varepsilon^j.
\end{eqnarray}
By the different choice of operator $L$,  Blaszak and Szum proposed 2+1 dimensional generalizations of some known lattice systems as well as some new lattice-field systems. It is well known that discrete zero curvature representation and infinitely many conservation laws are very important indicator of integrability of a lattice hierarchy. In this paper, we would like to investigate further the 2+1 dimensional lattice systems proposed by Blaszak and Szum. We first describe their discrete zero curvature representations. Then, by means of solving the corresponding discrete spectral equation, we demonstrate the existence of infinitely many conservation laws for them and derive the corresponding conserved densities and the associated fluxes formulaically. So, the integrability for these 2+1 dimensional lattice systems is further confirmed.
\setcounter{equation}{0}
\section{Discrete zero curvature representation for several 2+1 dimensional lattice hierarchies}
Considering the following spectral problem
\begin{eqnarray}
E\psi(n,t,y, \lambda)=U(n,t,y,\lambda)\psi(n,t,y,\lambda),\nonumber\\
\frac{d\psi(n,t,y, \lambda)}{dt}=V(n,t,y,\lambda)\psi(n,t,y,\lambda).
\end{eqnarray}
It is obvious that the condition of integrability of equation (2.1) leads to the following
discrete zero curvature representation:
\begin{eqnarray}
\frac{\partial {U}}{\partial t}=(EV)U-UV.
\end{eqnarray}
The lattice-field system derived from equation (2.2) is the Lax integrable. The matrix $U$ and $V$
are the Lax pairs of the system. In general, it is difficult to derive a 2+1 dimensional Lax integrable lattice-field system from equation (2.2), because we do not have a general method to construct proper matrix $U$ and $V$. For 2+1 dimensional lattice-field system (1.5), we note that it admits the following Lax form
\begin{eqnarray}
{L}_{t_i}=[A_i,{L}],
\end{eqnarray}
where
\begin{eqnarray*}
{L}=\tilde{L}-\partial_y, \qquad A_i=P_{\geq 0}\bigtriangledown C_i. 
\end{eqnarray*}
Therefore, we can obtain its discrete zero curvature representation. In the following, we illustrate in detail the discrete zero curvature representations for some 2+1 dimensional lattice hierarchies.\\
{\bf(1)} 2+1 dimensional Benjamin-Ono (B-O) lattice-field equation.\\
The Lax operator $L$ has the form
\begin{eqnarray}
L=\varepsilon+u-\partial_y.
\end{eqnarray}
Constructing $A_2$, $A_3$ et al as the follows:
\begin{eqnarray}
\lefteqn{}
&&A_2=\varepsilon^2+[u(n+1)+u(n)]\varepsilon+u^2(n)+Hu(n)_y,\nonumber\\
&&A_3=\varepsilon^3+[u(n+2)+u(n+1)+u(n)]\varepsilon^2+[u^2(n+1)+u(n+1)u(n)+u^2(n)\\ 
&&+\frac{1}{4}(E-1)(u(n)_y
+3H^2u(n)_y)]\varepsilon+\frac{1}{4}u(n)_{yy}\nonumber\\
&&+u(n)^3+\frac{3}{2}Hu(n)u(n)_y
+\frac{3}{2}u(n)Hu(n)_y+\frac{3}{4}H^2u(n)_{yy},\nonumber
\end{eqnarray}
$$\vdots$$
Then, the 2+1 dimensional B-O lattice hierarchy is presented by
\begin{eqnarray}
u(n)_{t_2}=2u(n)u(n)_y+Hu(n)_{yy},
\end{eqnarray}
\begin{eqnarray}
u(n)_{t_3}=\frac{1}{4}u(n)_{yyy}+3u(n)^2u(n)_y+\frac{3}{2}H(u(n)u(n)_y)_y+\frac{3}{2}(u(n)Hu(n)_y)_y+
\frac{3}{4}H^2u(n)_{yyy}
\end{eqnarray}
$$\vdots$$ 
where operator $H=(E+1)(E-1)^{-1}$, and $(E-1)^{-1}=-\sum_{k=0}^{\infty}E^k$.
The Lax pairs $U$ and $V$ for equations (2.6) and (2.7) are described by, respectively,
\begin{eqnarray}
\lefteqn{}
&&U=\lambda-u(n)+\partial_y,\\
&&V_2=(\lambda+\partial_y)^2-u(n)_y+Hu(n)_y,\\
&&V_3=(\lambda+\partial_y)(\lambda+\partial_y-u(n+1))(\lambda+\partial_y-u(n))\nonumber\\
&&+[u(n+1)+u(n)](\lambda+\partial_y-u(n+1))(\lambda+\partial_y-u(n))\nonumber\\
&&+[u^2(n+1)+u(n+1)u(n)+u^2(n) 
+\frac{1}{4}(E-1)(u(n)_y
+3H^2u(n)_y)](\lambda+\partial_y-u(n))\nonumber\\
&&+\frac{1}{4}u(n)_{yy}+u(n)^3+\frac{3}{2}Hu(n)u(n)_y
+\frac{3}{2}u(n)Hu(n)_y+\frac{3}{4}H^2u(n)_{yy}.
\end{eqnarray}
{\bf (2)} 2+1 dimensional Toda lattice-field equation.\\
The Lax operator $L$ is
\begin{eqnarray}
L=\varepsilon+p(n)+v(n)\varepsilon^{-1}-\partial_y.
\end{eqnarray}
Constructing $A_1$, $A_2$ et al as the follows:
\begin{eqnarray}
\lefteqn{}
&&A_1=\varepsilon+p(n),\nonumber\\
&&A_2=\varepsilon^2+[p(n)+p(n+1)]\varepsilon+p^2(n)+v(n)+v(n+1)+Hp(n)_y
\end{eqnarray}
$$\vdots$$
Then, the 2+1 dimensional Toda lattice hierarchy is written by
\begin{eqnarray}
\left(
\begin{array}{c}
v(n)\\
p(n)
\end{array}
\right)_{t_1}=\left(
\begin{array}{c}
v(n)[p(n)-p(n-1)]\\
v(n+1)-v(n)+p(n)_y
\end{array}
\right),
\end{eqnarray}
\begin{eqnarray}
\lefteqn{}\nonumber\\
&&v(n)_{t_2}=v(n)[p^2(n)-p^2(n-1)+v(n+1)-v(n-1)+p(n)_y+p(n-1)_y],\nonumber\\
&&p(n)_{t_2}=v(n+1)[p(n+1)+p(n)]-v(n)[p(n-1)+p(n)]+v(n)_y\nonumber\\
&&+v(n+1)_y+Hp(n)_{yy}+2p(n)p(n)_y
\end{eqnarray}
$$\vdots$$ 
Matrix Lax pairs $U$ and $V$ for equations (2.13) and (2.14) read as follows, respectively,
\begin{eqnarray}
U=\left(
\begin{array}{cc}
0&1\\
-v(n)&\lambda-p(n)+\partial_y,
\end{array}
\right)
\end{eqnarray}
\begin{eqnarray}
V_1=\left(
\begin{array}{cc}
p(n-1)&1\\
-v(n)&\lambda+\partial_y
\end{array}
\right)
\end{eqnarray}
\begin{eqnarray}
V_2=\left(
\begin{array}{cc}
v(n-1)+p^2(n-1)+Hp(n-1)_y&\lambda+\partial_y+p(n-1)\\
-v(n)(\lambda+\partial_y+p(n))-v(n)_y&(\lambda+\partial_y)^2-p(n)_y+v(n)+Hp(n)_y
\end{array}
\right)
\end{eqnarray}
{\bf (3)} The Lax operator $L$ is of the form
\begin{eqnarray}
L=\varepsilon^2+u(n)\varepsilon+v(n)-\partial_y.
\end{eqnarray}
Constructing $A_1$ as the follows:
\begin{eqnarray}
A_1=\varepsilon+(E+1)^{-1}u(n).
\end{eqnarray}
The first equation corresponding to the lattice hierarchy is
\begin{eqnarray}
\left(
\begin{array}{c}
u(n)\\
v(n)
\end{array}
\right)_{t_1}=\left(
\begin{array}{c}
v(n+1)-v(n)-u(n)H^{-1}u(n)\\
(E+1)^{-1}u(n)_y
\end{array}
\right)
\end{eqnarray}
Matrix Lax pairs $U$ and $V$ for equation (2.20) are presented by
\begin{eqnarray}
U=\left(
\begin{array}{cc}
0&1\\
\lambda-v(n-1)+\partial_y&-u(n-1)
\end{array}
\right)\\
V_1=\left(
\begin{array}{cc}
(E+1)^{-1}u(n-1)&1\\
\lambda-v(n-1)+\partial_y&-(E+1)^{-1}u(n-1)
\end{array}
\right)
\end{eqnarray}
{\bf (4)} 2+1 dimensional Blaszak-Szum (B-S) three-field lattice equation.\\  
The Lax operator $L$ takes the form
\begin{eqnarray}
L=\varepsilon^2+p(n)\varepsilon+v(n)+u(n)\varepsilon^{-1}-\partial_y.
\end{eqnarray}
Constructing $A_1$, $A_2$, $A_4$ et al as the follows:
\begin{eqnarray*}
A_1=\varepsilon+(E+1)^{-1}p(n),\\
A_2=\varepsilon^2+p(n)\varepsilon+v(n),
\end{eqnarray*}
\begin{eqnarray}
\lefteqn{}
&&A_4=\varepsilon^4+(p(n)+p(n+2))\varepsilon^3+(p(n)p(n+1)+v(n)+v(n+2))\varepsilon^2\nonumber\\
&&+[p(n)(v(n)+v(n+1))+u(n)+u(n+2)+(E^2-1)^{-1}(p(n)+p(n+2))_y]\varepsilon\nonumber\\
&&+v^2(n)+p(n-1)u(n)+p(n)u(n+1)+(E^2-1)^{-1}[p(n+1)(E^2-1)^{-1}(p(n)+p(n+2))_y\nonumber\\
&&-p(n)(E^2-1)^{-1}(p(n+1)+p(n+3))_y+(p(n)p(n+1)+v(n)+v(n+2))_y].
\end{eqnarray}
$$\vdots$$ 
Then, the first equations from the lattice hierarchy are
\begin{eqnarray}
\left(
\begin{array}{c}
u(n)\\
v(n)\\
p(n)
\end{array}
\right)_{t_1}=\left(
\begin{array}{c}
u(n)H^{-1}p(n-1)\\
u(n+1)-u(n)+(E+1)^{-1}p(n)_y\\
v(n+1)-v(n)-p(n)H^{-1}p(n)
\end{array}
\right),\\
\left(
\begin{array}{c}
u(n)\\
v(n)\\
p(n)
\end{array}
\right)_{t_2}=\left(
\begin{array}{c}
u(n)[v(n)-v(n-1)]\\
p(n)u(n+1)-p(n-1)u(n)+v(n)_y\\
u(n+2)-u(n)+p(n)_y
\end{array}
\right)
\end{eqnarray}
\begin{eqnarray}
\lefteqn{}
&&p(n)_{t_4}=(E^2-1)u(n)(v(n)+v(n-1))-p(n)(E+1)^{-1}[p(n+1)(E^2-1)^{-1}(p(n)+p(n+2))_y\nonumber\\
&&-p(n)(E^2-1)^{-1}(p(n+1)+p(n+3))_y+(p(n)p(n+1)+v(n)+v(n+2))_y]\nonumber\\
&&+(v(n+1)-v(n))(E^2-1)^{-1}(p(n)+p(n+2))_y\nonumber\\
&&+[p(n)(v(n)+v(n+1))+u(n)+u(n+2)+(E^2-1)^{-1}(p(n)+p(n+2))_y]_y\nonumber\\
&&u(n)_{t_4}=u(n)(1-E^{-1})[v^2(n)+p(n-1)u(n)+p(n)u(n+1)\nonumber\\
&&+(E^2-1)^{-1}(p(n+1)(E^2-1)^{-1}(p(n)+p(n+2))_y\nonumber\\
&&-p(n)(E^2-1)^{-1}(p(n+1)+p(n+3))_y+(p(n)p(n+1)+v(n)+v(n+2))_y)]\nonumber\\
&&v(n)_{t_4}=(1-E^{-1})u(n+1)[p(n)(v(n)+v(n+1))+u(n)+u(n+2)\nonumber\\
&&+(E^2-1)^{-1}(p(n)+p(n+2))_y]
+\partial_y[v^2(n)+p(n-1)u(n)+p(n)u(n+1)\nonumber\\
&&+(E^2-1)^{-1}[p(n+1)(E^2-1)^{-1}(p(n)+p(n+2))_y\nonumber\\
&&-p(n)(E^2-1)^{-1}(p(n+1)+p(n+3))_y+(p(n)p(n+1)+v(n)+v(n+2))_y]].
\end{eqnarray}
Matrix Lax pairs $U$ and $V$ for equations (2.25)-(2.27) read as follows, respectively,
\begin{eqnarray}
U=\left(
\begin{array}{ccc}
0&1&0\\
\lambda-v(n-1)+\partial_y&-p(n-1)&-u(n-1)\\
1&0&0
\end{array}
\right)
\end{eqnarray}
\begin{eqnarray}
V_1=\left(
\begin{array}{ccc}
(E+1)^{-1}p(n-1)&1&0\\
\lambda-v(n-1)+\partial_y&-(E+1)^{-1}p(n-1)&-u(n-1)\\
1&0&(E+1)^{-1}p(n-2)
\end{array}
\right)
\end{eqnarray}
\begin{eqnarray}
V_2=\left(
\begin{array}{ccc}
\lambda+\partial_y&0&-u(n-1)\\
-u(n)&\lambda+\partial_y&0\\
p(n-2)&1&v(n-2)
\end{array}
\right)
\end{eqnarray}
\begin{eqnarray}
V_4=(v_{ij})_{3\times 3}
\end{eqnarray}
where
\begin{eqnarray*}
\lefteqn{}\\
&&v_{11}=E^{-1}v_{22}+p_{n-2}(u_{n-1}+2(E^2-1)^{-1}p_{n-1,y}),\\
&&v_{12}=u_{n-1}+2(E^2-1)^{-1}p_{n-1,y},\\
&&v_{13}=-u_{n-1}(\lambda+\partial_y-v_{n-1})-u_{n-1,y},\\
&&v_{21}=-u_n(v_n+v_{n-1})+2((E^2-1)^{-1}p_{n,y})(\lambda+\partial_y-v_{n-1})-u_{n,y},\\ 
&&v_{22}=(\lambda+\partial_y)^2-v_{n,y}-2p_{n-1}(E^2-1)^{-1}p_{n,y}+(E^2-1)^{-1}[p_{n+1}(E^2-1)^{-1}(p_n+p_{n+2})_y\\
&&-p_n(E^2-1)^{-1}(p_{n+1}+p_{n+3})_y+(p_np_{n+1}+v_n+v_{n+2})_y],\\
&&v_{23}=-u_{n-1}(u_n+2(E^2-1)^{-1}p_{n,y}),\\
&&v_{31}=u_{n-2}+2(E^2-1)^{-1}p_{n-2,y}+p_{n-2,y}+p_{n-2}(\lambda+\partial_y-v_{n-2}),\\
&&v_{32}=\lambda+\partial_y-v_{n-2},\\
&&v_{33}=p_{n-3}u_{n-2}+2v_{n-2}(\lambda+\partial_y)-v_{n-2}^2+(E^2-1)^{-1}[p_{n-1}(E^2-1)^{-1}(p_n+p_{n-2})_y\\
&&-p_{n-2}(E^2-1)^{-1}(p_{n+1}+p_{n-1})_y+(p_{n-2}p_{n-1}+v_{n-2}+v_{n})_y],\\
\end{eqnarray*}
{\bf (5)} 2+1 dimensional B-S four-field lattice equation.\\
The Lax operator $L$ reads
\begin{eqnarray}
L=\varepsilon^3+p(n)\varepsilon^2+v(n)\varepsilon+u(n)+q(n)\varepsilon^{-1}-\partial_y.
\end{eqnarray}
Constructing $A_1$, $A_2$, $A_3$ et al as the follows:
\begin{eqnarray}
\lefteqn{}
&&A_1=\varepsilon+(E^2+E+1)^{-1}p(n),\nonumber\\
&&A_2=\varepsilon^2+(E^2+E+1)^{-1}(E+1)p(n)\varepsilon+\nonumber\\
&&(E^2+E+1)^{-1}[v(n+1)+v(n)-p(n)(E^2+E+1)^{-1}p(n+1)],\\
&&A_3=\varepsilon^3+p(n)\varepsilon^2+v(n)\varepsilon+u(n)\nonumber
\end{eqnarray}
$$\vdots$$
where operator $(E^2+E+1)^{-1}$ is defined by
\begin{eqnarray*}
(E^2+E+1)^{-1}=\sum_{k=0}^{\infty}(E^{3k}-E^{3k+1})
\end{eqnarray*}
Then, the first equations from the lattice hierarchy are written by the following equations:
\begin{eqnarray}
\left(
\begin{array}{c}
u(n)\\
v(n)\\
p(n)\\
q(n)
\end{array}
\right)_{t_1}=\left(
\begin{array}{c}
q(n+1)-q(n)+(E^2+E+1)^{-1}p(n)_y\\
u(n+1)-u(n)+v(n)(1-E)(E^2+E+1)^{-1}p(n)\\
v(n+1)-v(n)-p(n)(1-E^2)(E^2+E+1)^{-1}p(n)\\
q(n)(E-1)(E^2+E+1)^{-1}p(n-1)
\end{array}
\right),
\end{eqnarray}
\begin{eqnarray}
\left(
\begin{array}{c}
u(n)\\
v(n)\\
p(n)\\
q(n)
\end{array}
\right)_{t_2}=\left(
\begin{array}{c}
\Delta_1\\
\Delta_2\\
\Delta_3\\
\Delta_4\\
\end{array}
\right),
\end{eqnarray}
where
\begin{eqnarray*}
\lefteqn{}\\
&&\Delta_1=(E-1)[q(n)(E^2+E+1)^{-1}(p(n)+p(n-1))]\\
&&+\partial_y[(E^2+E+1)^{-1}(v(n)+v(n+1)-p(n)(E^2+E+1)^{-1}p(n+1))]\\
&&\Delta_2=q(n+2)-q(n)+(u(n+1)-u(n))(E^2+E+1)^{-1}(E+1)p(n)\\
&&-v(n)(E-1)(E^2+E+1)^{-1}[v(n+1)+v(n)-p(n)(E^2+E+1)^{-1}p(n+1)]\\
&&+(E^2+E+1)^{-1}(p(n)+p(n+1))_y\\
&&\Delta_3=u(n+2)-u(n)-v(n+1)(E^2+E+1)^{-1}p(n+2)+\\
&&v(n)(E^2+E+1)^{-1}p(n)-p(n)(E^2+E+1)^{-1}(v_{n+2}-v_{n+1})\\
&&+p_n(E^2-1)(E^2+E+1)^{-1}[p(n)(E^2+E+1)^{-1}p(n+1)]\\
&&\Delta_4=q(n)(E-1)(E^2+E+1)^{-1}[v(n)+v(n-1)-p(n-1)(E^2+E+1)^{-1}p(n)]
\end{eqnarray*}
\begin{eqnarray}
\left(
\begin{array}{c}
u(n)\\
v(n)\\
p(n)\\
q(n)
\end{array}
\right)_{t_3}=\left(
\begin{array}{c}
v(n)q(n+1)-v(n-1)q(n)+u(n)_y\\
p(n)q(n+2)-p(n-1)q(n)+v(n)_y\\
q(n+3)-q(n)+p(n)_y\\
q(n)(u(n)-u_{n-1})
\end{array}
\right)
\end{eqnarray}
Matrix Lax pairs $U$ and $V$ for equations (2.34)-(2.36) are, respectively,
\begin{eqnarray}
U=\left(
\begin{array}{cccc}
0&1&0&0\\
0&0&1&0\\
\lambda-u(n-2)+\partial_y&-v(n-2)&-p(n-2)&-q(n-2)\\
1&0&0&0
\end{array}
\right)
\end{eqnarray}
\begin{eqnarray}
\small{
V_1=\left(
\begin{array}{cccc}
(E^2+E+1)^{-1}p_{n-2}&1&0&0\\
0&(E^2+E+1)^{-1}p_{n-1}&1&0\\
\lambda+\partial_y-u_{n-2}&-v_{n-2}&(E^2+E+1)^{-1}p_n-p_{n-2}&-q_{n-2}\\
1&0&0&(E^2+E+1)^{-1}p_{n-3}
\end{array}
\right)
}
\end{eqnarray}
\begin{eqnarray}
V_2=(v_{ij})_{4\times 4},
\end{eqnarray}
where
\begin{eqnarray*}
\lefteqn{}\\
&&v_{11}=(E^2+E+1)^{-1}[v_{n-2}+v_{n-1}-p_{n-2}(E^2+E+1)^{-1}p_{n-1}],\\
&&v_{12}=(E^2+E+1)^{-1}(p_{n-1}+p_{n-2}), \qquad v_{13}=1, \qquad v_{14}=0,\\
&&v_{21}=\lambda+\partial_y-u_{n-2}, \qquad v_{22}=-(E^2+E+1)^{-1}[v_{n-2}+p_{n-1}(E^2+E+1)^{-1}p_n],\\
&&v_{23}=-(E^2+E+1)^{-1}p_{n-2}, \qquad v_{24}=-q_{n-2},\\
&&v_{31}=[(E^2+E+1)^{-1}p_{n-1}](u_{n-2}-\lambda-\partial_y)-q_{n-1},\\
&&v_{32}=\lambda+\partial_y-u_{n-1}+v_{n-2}(E^2+E+1)^{-1}p_{n-1}\\
&&v_{33}=
p_{n-2}(E^2+E+1)^{-1}p_{n-1}-(E^2+E+1)^{-1}[v_{n-1}+p_n(E^2+E+1)^{-1}p_{n+1}],\\
&&v_{34}=q_{n-2}(E^2+E+1)^{-1}p_{n-1}\\
&&v_{41}=(E^2+E+1)^{-1}(p_{n-2}+p_{n-3}), \qquad v_{42}=1, \qquad v_{43}=0,\\
&&v_{44}=(E^2+E+1)^{-1}[v_{n-3}+v_{n-2}-p_{n-3}(E^2+E+1)^{-1}p_{n-2}],
\end{eqnarray*}
\begin{eqnarray}
V_3=\left(
\begin{array}{cccc}
\lambda+\partial_y&0&0&-q(n-2)\\
-q(n-1)&\lambda+\partial_y&0&0\\
0&-q(n)&\lambda+\partial_y&0\\
v(n-3)&p(n-3)&1&u(n-3)
\end{array}
\right)
\end{eqnarray}
\setcounter{section}{2}
\setcounter{equation}{0}
\section{Infinitely many conservation laws for several 2+1 dimensional lattice hierarchies}
In this section, we show that the 2+1 dimensional lattice hierarchies described in section 1
possess infinitely many conservation laws
and the corresponding conserved densities and the associated fluxes are derived formulaically. 
Here the important procedure is solving the corresponding discrete spectral equation
\begin{eqnarray}
L \psi_n=\lambda \psi_n.
\end{eqnarray}
{\bf (1)}. Infinitely many conservation laws for 2+1 dimensional Benjamin-Ono lattice-field equation\\ 
The 2+1 dimensional Benjamin-Ono lattice-field equations correspond to the discrete spectral problem 
\begin{eqnarray}
\frac{\psi_{n+1}}{\psi_n}=\lambda-u_n+\partial_yln\psi_n
\end{eqnarray}
Let
$\Gamma_n=\frac{\psi_{n+1}}{\psi_{n}},$ the spectral problem leads to
\begin{eqnarray}
\Gamma_n\Gamma_{n+1}-\Gamma_n^2+(u_{n+1}-u_n)\Gamma_n=\frac{\partial \Gamma_n}{\partial y}
\end{eqnarray}
Suppose the eigenfunctions $\psi_n$
is an analytical function of the arguments and expand $\Gamma_n$ with respect to $\lambda$ by the Taylor series
\begin{eqnarray}
\Gamma_n=\sum_{j=1}^{\infty}\lambda^{-j}w_n^{(j)},
\end{eqnarray}
and substitute equation (3.4) into equation (3.3), we obtain
\begin{eqnarray*}
\lefteqn{}
&&w_n^{(1)}=exp(\int_{0}^{y}(u_{n+1}-u_n)dy),\nonumber\\ 
&&w_n^{(2)}=w_n^{(1)}[1+\int_{0}^{y}(E-1)exp(\int_{0}^{y}(u_{n+1}-u_n)dy)dy],\nonumber\\
&&w_n^{(i)}=w_n^{(1)}[1+\int_{0}^{y}exp(\int_{0}^{y}(u_{n}-u_{n+1})dy)\sum_{l+s=i}w_n^{(l)}(w_{n+1}^{(s)}-w_n^{(s)})dy], i\geq 3.
\end{eqnarray*}
On the other hand, it is obvious that
\begin{eqnarray}
\frac{\partial}{\partial t}\ln\Gamma_n=(E-1)\frac{\partial}{\partial t}\ln\psi_n.
\end{eqnarray}
For equations (2.6) and (2.7), $\frac{\partial}{\partial t}ln \psi_n$ is described by the following equations, respectively,
\begin{eqnarray}
\frac{\partial}{\partial t}\ln\psi_n=u_n^2+Hu_{n,y}+2u_n\Gamma_n+\Gamma_n^2+\frac{\partial \Gamma_{n}}{\partial y}
\end{eqnarray}
and 
\begin{eqnarray}
\frac{\partial}{\partial t}\ln\psi_n=
\frac{1}{4}u(n)_{yy}+u(n)^3+\frac{3}{2}Hu(n)u(n)_y
+\frac{3}{2}u(n)Hu(n)_y+\frac{3}{4}H^2u(n)_{yy}\nonumber\\
+\Gamma_n^3+3u_n^2\Gamma_n+3u_n\Gamma_n^2+\frac{3}{2}(Hu_{n,y}-u_{n,y})\Gamma_n
+\frac{\partial}{\partial y}(\Gamma_n\Gamma_{n+1}+\frac{1}{2}\Gamma_n^2+2u_n\Gamma_n+u_{n+1}\Gamma_n)
\end{eqnarray}
It follows from equation (3.5) that
\begin{eqnarray}
\frac{\partial}{\partial t}ln \Gamma_n+\frac{\partial}{\partial y}(1-E)\Gamma_n=(E-1)(u_n^2+Hu_{n,y}+2u_n\Gamma_n+\Gamma_n^2)
\end{eqnarray}
and
\begin{eqnarray}
\lefteqn{}
&&\frac{\partial}{\partial t}ln \Gamma_n+\frac{\partial}{\partial y}(1-E)[\Gamma_n\Gamma_{n+1}+\frac{1}{2}\Gamma_n^2+(2u_n+u_{n+1})\Gamma_n]\nonumber\\
&&=(E-1)[\frac{1}{4}u(n)_{yy}+u(n)^3+\frac{3}{2}Hu(n)u(n)_y\nonumber\\
&&+\frac{3}{2}u(n)Hu(n)_y+\frac{3}{4}H^2u(n)_{yy}
+\Gamma_n^3+3u_n\Gamma_n^2+(3u_n^2+\frac{3}{2}Hu_{n,y}-\frac{3}{2}u_{n,y})\Gamma_n]
\end{eqnarray}
Note that
\begin{eqnarray}
\frac{\partial}{\partial t}ln \Gamma_n=\frac{\partial}{\partial t}\ln w_n^{(1)}+\frac{\partial}{\partial t}(\sum_{k=1}^{\infty}\frac{(-1)^{k+1}\lambda^{-k}}{k}\Phi^k)
\end{eqnarray}
where
\begin{eqnarray}
\Phi=\sum_{j=0}^{\infty}\lambda^{-j}s_j, \qquad s_j=\bar{w}_n^{(j+2)}=\frac{{w}_n^{(j+2)}}{w_n^{(1)}}, 
\end{eqnarray}
Making a comparison of the powers of $\lambda$ on both sides of equations (3.8) and (3.9), infinitely many conservation laws of the 2+1 dimension B-O lattice-field equations (2.6) and (2.7) are obtained,
\begin{eqnarray}
\frac{\partial}{\partial t}\rho_{n}^{(j)}+\frac{\partial}{\partial y}\beta_{n}^{(j)}=(E-1)J_n^{(j)},\qquad
j=0,1,2,.....
\end{eqnarray}
For equation (2.6), the conserved densities and the associated fluxes are 
\begin{eqnarray}
\rho_{n}^{(0)}=\int_{0}^{y}(u_{n+1}-u_n)dy,\qquad \beta_{n}^{(0)}=0,\qquad J_n^{(0)}=u_n^2+Hu_{n,y},\nonumber\\
\rho_{n}^{(1)}=\bar{w}_n^{(2)},\qquad \beta_{n}^{(1)}=(1-E)w_n^{(1)}, \qquad J_n^{(1)}=2u_nw_n^{(1)}
\end{eqnarray}
and
\begin{eqnarray}
\rho_n^{(j)}=s_{j-1}-\frac{1}{2}\sum_{l_1+l_2=j-2}s_{l_1}s_{l_2}+\frac{1}{3}\sum_{l_1+l_2+l_3=j-3}s_{l_1}s_{l_2}s_{l_3}\nonumber\\
-.....+\frac{(-1)^{j-1}}{j-2}\sum_{l_1+l_2+...+l_{j-2}=2}s_{l_1}s_{l_2}...s_{l_{j-2}}+
(-1)^js_0^{j-2}s_1+\frac{(-1)^{j+1}}{j}s_0^j,
\end{eqnarray}
\begin{eqnarray}
\beta_n^{(j)}=(1-E)w_n^{(j)}, \qquad J_n^{(j)}=2u_nw_n^{(j)}+\sum_{l+s=j}w_n^{(l)}w_n^{(s)},
\qquad j=2,3,.....
\end{eqnarray}
For equation (2.7), we have
\begin{eqnarray}
\lefteqn{}
&&\rho_{n}^{(0)}=\int_{0}^{y}(u_{n+1}-u_n)dy,\qquad \beta_{n}^{(0)}=0,\nonumber\\ &&J_n^{(0)}=\frac{1}{4}u(n)_{yy}+u(n)^3+\frac{3}{2}Hu(n)u(n)_y
+\frac{3}{2}u(n)Hu(n)_y+\frac{3}{4}H^2u(n)_{yy},\nonumber\\
&&\rho_{n}^{(1)}=\bar{w}_n^{(2)},\qquad \beta_{n}^{(1)}=(1-E)[(2u_n+u_{n+1})w_n^{(1)}],
\nonumber\\ 
&&J_n^{(1)}=(3u_n^2+\frac{3}{2}Hu_{n,y}-\frac{3}{2}u_{n,y})w_n^{(1)},
\end{eqnarray}
and
$\rho_n^{(j)}$ (j=2,3,...) is presented by equation (3.14) and $\beta_n^{(j)}$ and $J_n^{(j)}$
are 
\begin{eqnarray}
\lefteqn{}
&&\beta_n^{(j)}=(1-E)[(2u_n+u_{n+1})w_n^{(j)}+\sum_{l+s=j}w_n^{(l)}(\frac{w_n^{(s)}}{2}+w_{n+1}^{(s)})]\nonumber\\
&&J_n^{(j)}=(3u_n^2+\frac{3}{2}Hu_{n,y}-\frac{3}{2}u_{n,y})w_n^{(j)}\nonumber\\
&&+3u_n\sum_{l+s=j}w_n^{(l)}w_n^{(s)}+\sum_{l+s+m=j}w_n^{(l)}w_n^{(s)}w_n^{(m)},
\qquad j=2,3,....
\end{eqnarray}
{\bf (2)}. Infinitely many conservation laws for 2+1 dimensional Toda lattice-field equation\\ 
The 2+1 dimensional Toda lattice-field equation corresponds to the discrete spectral problem 
\begin{eqnarray}
\frac{\psi_{n+1}}{\psi_n}=\lambda-p_n-\frac{v_n\psi_{n-1}}{\psi_n}+\frac{\partial}{\partial y}ln\psi_n,
\end{eqnarray}
i.e.,
\begin{eqnarray}
\Gamma_{n-1}\Gamma_n(\Gamma_{n+1}-\Gamma_n+p_{n+1}-p_n)+v_{n+1}\Gamma_{n-1}-v_n\Gamma_n=\Gamma_{n-1}\frac{\partial \Gamma_n}{\partial y},
\end{eqnarray}
where $\Gamma_n=\frac{\psi_{n+1}}{\psi_{n}}.$
Suppose the eigenfunctions $\psi_n$
is an analytical function of the arguments and expand $\Gamma_n$ with respect to $\lambda$ by the Taylor series (3.4)
and substitute it into equation (3.19), we have
\begin{eqnarray*}
\lefteqn{}
&&w_n^{(1)}=v_{n+1},\nonumber\\ 
&&w_n^{(2)}=v_{n+1}[p_{n+1}-(E-1)^{-1}\partial_y\ln v_{n+2}],\nonumber\\
&&w_n^{(i)}=v_{n+1}(E-1)^{-1}[\frac{A(n)}{v_{n+1}v_{n+2}}],\qquad i\geq 3\nonumber
\end{eqnarray*}
where
\begin{eqnarray*}
A(n)=(p_{n+2}-p_{n+1})\sum_{l+s=i}w_n^{(l)}w_{n+1}^{(s)}+
&&\sum_{l+s+m=i}w_n^{(l)}w_{n+1}^{(s)}(w_{n+2}^{(m)}-w_{n+1}^{(m)})-\sum_{l+s=i}w_n^{(l)}w_{n+1,y}^{(s)}.
\end{eqnarray*}
For 2+1 dimensional Toda lattice-field equations (2.13) and (2.14), $\frac{\partial}{\partial t}ln \psi_n$ is written in the form, respectively,
\begin{eqnarray}
\frac{\partial}{\partial t}\ln\psi_n
=p_n+\Gamma_n
\end{eqnarray}
and 
\begin{eqnarray}
\frac{\partial}{\partial t}\ln\psi_n
=v_n+p_n^2+Hp_{n,y}+\Gamma_{n,y}+2p_n\Gamma_n+\frac{v_n\Gamma_n}{\Gamma_{n-1}}+\Gamma_n^2\nonumber\\
=v_n+v_{n+1}+p_n^2+Hp_{n,y}+(p_n+p_{n+1})\Gamma_{n}+\Gamma_n\Gamma_{n+1}
\end{eqnarray}
By means of equation (3.5), we obtain the following discrete conservation equations:
\begin{eqnarray}
\frac{\partial}{\partial t}ln \Gamma_n=(E-1)(p_n+\Gamma_n)
\end{eqnarray}
and
\begin{eqnarray}
\frac{\partial}{\partial t}ln \Gamma_n
=(E-1)[v_n+v_{n+1}+p_n^2+Hp_{n,y}+(p_n+p_{n+1})\Gamma_{n}+\Gamma_n\Gamma_{n+1}
]
\end{eqnarray}
In comparison with the powers of $\lambda$ on both sides of equation (3.22) and (3.23), infinitely many conservation laws of the 2+1 dimension Toda lattice-field equations (2.13) and (2.14) are derived,
\begin{eqnarray}
\frac{\partial}{\partial t}\rho_{n}^{(j)}=(E-1)J_n^{(j)},\qquad
j=0,1,2,.....
\end{eqnarray}
Here, the conserved densities possess the form
\begin{eqnarray*}
\rho_{n}^{(0)}=\ln v_{n+1},\qquad \rho_{n}^{(1)}=p_{n+1}-(E-1)^{-1}\partial_yln v_{n+2},
\end{eqnarray*}
and $\rho_n^{(j)},j\geq 2$ is written by equation (3.14). The associated fluxes for equations (2.13) and (2.14) are described by the following, respectively,
\begin{eqnarray}
\lefteqn{}
&&J_n^{(0)}=p_n,\qquad J_n^{(1)}=v_{n+1}, \qquad J_n^{(j)}=w_n^{(j)}, j\geq 2\nonumber\\
&&J_n^{(0)}=v_n+v_{n+1}+p_n^2+Hp_{n,y},\qquad 
J_n^{(1)}=(p_n+p_{n+1})v_{n+1},\nonumber\\
&&J_n^{(j)}=(p_n+p_{n+1})w_n^{(j)}+\sum_{l+s=j}w_n^{(l)}w_{n+1}^{(s)},\qquad j\geq 2
\end{eqnarray}
{\bf (3)}. Infinitely many conservation laws for 2+1 dimensional lattice-field equation (2.20)\\ 
Discrete spectral equation corresponding to the 2+1 dimensional lattice-field equation (2.20) 
is 
\begin{eqnarray}
\frac{\psi_{n+1}}{\psi_n}=\frac{(\lambda-v_{n-1}+\partial_y)\psi_{n-1}}{\psi_n}-u_{n-1}
\end{eqnarray}
i.e.,
\begin{eqnarray}
\Gamma_{n-1}(\Gamma_n\Gamma_{n+1}-\Gamma_{n-1}\Gamma_n+v_{n}-v_{n-1}+u_n\Gamma_{n}-u_{n-1}\Gamma_{n-1})=\frac{\partial \Gamma_{n-1}}{\partial y}
\end{eqnarray}
where $\Gamma_n=\frac{\psi_{n+1}}{\psi_{n}}.$ 
Suppose the eigenfunctions $\psi_n$
is an analytical function of the arguments and expand $\Gamma_n$ with respect to $\lambda$ by the Taylor series (3.4)
and substitute it into equation (3.27), we obtain
\begin{eqnarray*}
\lefteqn{}
&&w_n^{(1)}=exp(\int_{0}^{y}(v_{n+1}-v_n)dy),\nonumber\\ 
&&w_n^{(2)}=w_n^{(1)}[1+\int_{0}^{y}(E-1)u_nw_n^{(1)}dy],\nonumber\\
&&w_n^{(i)}=w_n^{(1)}[1+\int_{0}^{y}exp(\int_{0}^{y}(v_{n}-v_{n+1})dy)B(n)dy],\qquad i\geq 3,
\end{eqnarray*}
where
\begin{eqnarray*}
B(n)=u_{n+1}\sum_{l+s=i}w_n^{(l)}w_{n+1}^{(s)}-u_n\sum_{l+s=i}w_n^{(l)}w_{n}^{(s)}
-\sum_{l+s+m=i}w_n^{(l)}w_{n+1}^{(s)}(w_{n+2}^{(m)}-w_n^{(m)})
\end{eqnarray*}
Note that
\begin{eqnarray}
\frac{\partial}{\partial t}\ln\psi_n
=\Gamma_n+(E+1)^{-1}u_n.
\end{eqnarray}
From equation (3.5), we get the following discrete conservation equation:
\begin{eqnarray}
\frac{\partial}{\partial t}ln \Gamma_n=(E-1)(\Gamma_n+(E+1)^{-1}u_n)
\end{eqnarray}
Making a comparison of the powers of $\lambda$ on both sides of equation (3.29), infinitely many conservation laws of the 2+1 dimension lattice-field equations (2.20) are proposed,
\begin{eqnarray}
\frac{\partial}{\partial t}\rho_{n}^{(j)}=(E-1)J_n^{(j)},\qquad
j=0,1,2,.....
\end{eqnarray}
where
\begin{eqnarray}
\rho_{n}^{(0)}=\int_{0}^{y}(v_{n+1}-v_n)dy,\qquad J_n^{(0)}=(E+1)^{-1}u_n,\nonumber\\
\rho_{n}^{(1)}=\bar{w}_n^{(2)}, \qquad J_n^{(1)}=exp(\int_{0}^{y}(v_{n+1}-v_n)dy),\qquad 
J_n^{(j)}=w_n^{(j)}, j\geq 2.
\end{eqnarray}
and $\rho_n^{(j)}, j\geq 2$ is given by equation (3.14).\\ 
{\bf (4)}. Infinitely many conservation laws for 2+1 dimensional B-S three lattice-field equations\\ 
The 2+1 dimensional B-S three lattice-field equations (2.25) and (2.26) correspond to the discrete spectral problem 
\begin{eqnarray}
{\psi_{n+1}}+p_{n-1}\psi_n+v_{n-1}\psi_{n-1}+u_{n-1}\psi_{n-2}=(\lambda+\partial_y)\psi_{n-1},
\end{eqnarray}
which leads to the following discrete Riccati-type equation,
\begin{eqnarray}
\lefteqn{}
&&\Gamma_{n-1}\Gamma_{n}\Gamma_{n+1}(\Gamma_{n+2}-\Gamma_{n}+p_{n+1})\nonumber\\
&&+\Gamma_{n-1}\Gamma_n(v_{n+1}-v_n-p_n\Gamma_n)+u_{n+1}\Gamma_{n-1}-u_n\Gamma_n=\Gamma_{n-1}\frac{\partial \Gamma_{n}}{\partial y},
\end{eqnarray}
where
$\Gamma_n=\frac{\psi_{n+1}}{\psi_{n}}$.
Suppose the eigenfunctions $\psi_n$
is an analytical function of the arguments and expand $\Gamma_n$ with respect to $\lambda$ by the Taylor series (3.4)
and substitute it into equation (3.33), we have
\begin{eqnarray*}
\lefteqn{}
&&w_n^{(1)}=u_{n+1},\nonumber\\ 
&&w_n^{(2)}=u_{n+1}[v_{n+1}-(E-1)^{-1}\partial_y\ln u_{n+2}],\nonumber\\
&&w_n^{(i)}=-u_{n+1}(E-1)^{-1}[\frac{C(n)}{u_{n+1}u_{n+2}}]\nonumber\\
\end{eqnarray*}
where
\begin{eqnarray*}
\lefteqn{}
&&C(n)=\sum_{l+s=i}w_n^{(l)}w_{n+1,y}^{(s)}+
(v_{n+1}-v_{n+2})\sum_{l+s=i}w_n^{(l)}w_{n+1}^{(s)}\nonumber\\
&&+p_{n+1}\sum_{l+s+m=i}w_n^{(l)}w_{n+1}^{(s)}w_{n+1}^{(m)}
-p_{n+2}\sum_{l+s+m=i}w_n^{(l)}w_{n+1}^{(s)}w_{n+2}^{(m)}\nonumber\\
&&-\sum_{l+s+m+r=i}w_{n}^{(l)}w_{n+1}^{(s)}w_{n+2}^{(m)}(w_{n+3}^{(r)}-w_{n+1}^{(r)}), i=3,4,......\\
\end{eqnarray*}
Note that $\frac{\partial}{\partial t}\ln\psi_n$ for equations (2.25) and (2.26) are written by the following equations, respectively,
\begin{eqnarray}
\frac{\partial}{\partial t}\ln\psi_n
=\Gamma_n+(E+1)^{-1}p_n,
\end{eqnarray}
\begin{eqnarray}
\frac{\partial}{\partial t}\ln\psi_n
=\Gamma_{n}\Gamma_{n+1}+p_n\Gamma_n+v_n-\lambda,
\end{eqnarray}
we thus obtain two discrete conservation equations,
\begin{eqnarray}
\frac{\partial}{\partial t}ln \Gamma_n=(E-1)(\Gamma_n+(E+1)^{-1}p_n)
\end{eqnarray}
\begin{eqnarray}
\frac{\partial}{\partial t}ln \Gamma_n=(E-1)(\Gamma_{n}\Gamma_{n+1}+p_n\Gamma_n+v_n).
\end{eqnarray}
Making a comparison of the powers of $\lambda$ on both sides of equations (3.36) and (3.37), infinitely many conservation laws of the 2+1 dimension B-S three lattice-field equations (2.25) and (2.26) are proposed,
\begin{eqnarray}
\frac{\partial}{\partial t}\rho_{n}^{(j)}=(E-1)J_n^{(j)},\qquad
j=0,1,2,.....
\end{eqnarray}
where the conserved densities have the form
\begin{eqnarray*}
\rho_{n}^{(0)}=ln u_{n+1},\qquad \rho_{n}^{(1)}=v_{n+1}-(E-1)^{-1}\partial_yln u_{n+2},
\end{eqnarray*}
and $\rho_n^{(j)}, j\geq 2$ is described by equation (3.14). The
associated fluxes for equation (2.25) and (2.26) are given, respectively,
\begin{eqnarray}
J_n^{(0)}=(E+1)^{-1}p_n,\qquad J_n^{(1)}=u_{n+1},\qquad J_n^{(j)}=w_n^{(j)}, \qquad j\geq 2
\end{eqnarray}
and 
\begin{eqnarray}
J_n^{(0)}=v_n,\qquad J_n^{(1)}=p_nu_{n+1}, \qquad J_n^{(j)}=p_nw_n^{(j)}+\sum_{l+s=j}w_n^{(l)}w_{n+1}^{(s)},
\qquad j\geq 2
\end{eqnarray} 
{\bf (5)}. Infinitely many conservation laws for 2+1 dimensional B-S four lattice-field equations\\ 
Discrete spectral problem corresponding to
the 2+1 dimensional B-S four lattice-field equations (2.34)-(2.36) reads
\begin{eqnarray}
\psi_{n+1}+p_{n-2}\psi_n+v_{n-2}\psi_{n-1}+u_{n-2}\psi_{n-2}+q_{n-2}\psi_{n-3}=(\lambda+\partial_y)\psi_{n-2},
\end{eqnarray}
which leads to the discrete Riccati-type equation,
\begin{eqnarray}
\lefteqn{}
&&\Gamma_{n-1}\Gamma_{n}\Gamma_{n+1}\Gamma_{n+2}(\Gamma_{n+3}-\Gamma_{n}+p_{n+1})
+\Gamma_{n-1}\Gamma_n\Gamma_{n+1}(v_{n+1}-p_n\Gamma_n)\nonumber\\
&&+\Gamma_{n-1}\Gamma_{n}(u_{n+1}-u_n-v_n\Gamma_n)
+q_{n+1}\Gamma_{n-1}-q_n\Gamma_n=\Gamma_{n-1}\frac{\partial \Gamma_{n}}{\partial y}
\end{eqnarray}
where
$\Gamma_n=\frac{\psi_{n+1}}{\psi_{n}}$.
Suppose the eigenfunctions $\psi_n$
is an analytical function of the arguments and expand $\Gamma_n$ with respect to $\lambda$ by the Taylor series (3.4)
and substitute it into equation (3.42), we have
\begin{eqnarray}
\lefteqn{}
&&w_n^{(1)}=q_{n+1},\nonumber\\ 
&&w_n^{(2)}=q_{n+1}[u_{n+1}-(E-1)^{-1}\partial_y\ln q_{n+2}],\nonumber\\
&&w_n^{(i)}=-q_{n+1}(E-1)^{-1}[\frac{D(n)}{q_{n+1}q_{n+2}}],\qquad i\geq 3
\end{eqnarray}
where
{\small
\begin{eqnarray*}
\lefteqn{}
&&D(n)=\sum_{l+s=i}w_n^{(l)}w_{n+1,y}^{(s)}+
(u_{n+1}-u_{n+2})\sum_{l+s=i}w_n^{(l)}w_{n+1}^{(s)}+
(v_{n+1}-v_{n+2})\sum_{l+s+m=i}w_n^{(l)}w_{n+1}^{(s)}(w_{n+1}^{(m)}+w_{n+2}^{(m)})\nonumber\\
&&+(p_{n+1}-p_{n+2})\sum_{l+s+m+\mu=i}w_n^{(l)}w_{n+1}^{(s)}w_{n+2}^{(m)}(w_{n+1}^{(\mu)}+w_{n+3}^{(\mu)})
+\sum_{l+s+m+\mu+\gamma=i}w_{n}^{(l)}w_{n+1}^{(s)}w_{n+2}^{(m)}w_{n+3}^{(\mu)}(w_{n+1}^{(\gamma)}-w_{n+4}^{(\gamma)}).
\end{eqnarray*}
}
Note that $\frac{\partial}{\partial t}\ln\psi_n$ for equations (2.34)-(2.36) are written by the following equations, respectively,
\begin{eqnarray}
\frac{\partial}{\partial t}\ln\psi_n
=\Gamma_n+(E^2+E+1)^{-1}p_{n},
\end{eqnarray}
\begin{eqnarray}
\frac{\partial}{\partial t}\ln\psi_n
=\Gamma_n\Gamma_{n+1}+\Gamma_n(E^2+E+1)^{-1}(p_{n+1}+p_n)\nonumber\\
+(E^2+E+1)^{-1}[v_{n+1}+v_n-p_n(E^2+E+1)^{-1}p_{n+1}],
\end{eqnarray}
\begin{eqnarray}
\frac{\partial}{\partial t}\ln\psi_n
=u_n+v_n\Gamma_{n}+p_n\Gamma_n\Gamma_{n+1}+\Gamma_n\Gamma_{n+1}\Gamma_{n+2},
\end{eqnarray}
It follows from equation (3.5) that
\begin{eqnarray}
\frac{\partial}{\partial t}ln \Gamma_n=(E-1)[\Gamma_n+(E^2+E+1)^{-1}p_{n}]
\end{eqnarray}
\begin{eqnarray}
\frac{\partial}{\partial t}ln \Gamma_n=(E-1)[\Gamma_n\Gamma_{n+1}+\Gamma_n(E^2+E+1)^{-1}(p_{n+1}+p_n)\nonumber\\
+(E^2+E+1)^{-1}(v_{n+1}+v_n-p_n(E^2+E+1)^{-1}p_{n+1})],
\end{eqnarray}
\begin{eqnarray}
\frac{\partial}{\partial t}ln \Gamma_n=(E-1)(u_n+v_n\Gamma_{n}+p_n\Gamma_n\Gamma_{n+1}+\Gamma_n\Gamma_{n+1}\Gamma_{n+2}).
\end{eqnarray}
In comparison with the powers of $\lambda$ on both sides of equations (3.47)-(3.49), we obtain infinitely many conservation laws of the 2+1 dimension B-S four lattice-field equations (2.34)-(2.36),
\begin{eqnarray}
\frac{\partial}{\partial t}\rho_{n}^{(j)}=(E-1)J_n^{(j)}.\qquad
j=0,1,2,.....
\end{eqnarray}
where
\begin{eqnarray}
\rho_{n}^{(0)}=ln q_{n+1},\qquad \rho_{n}^{(1)}=u_{n+1}-(E-1)^{-1}\partial_y\ln q_{n+2},
\end{eqnarray}
and $\rho_n^{(j)}, j\geq 2$ is presented by equation (3.14). The corresponding fluxes for equations (2.34)-(2.36) are, respectively, 
\begin{eqnarray}
\qquad J_n^{(0)}=(E^2+E+1)^{-1}p_n,\qquad J_n^{(1)}=q_{n+1},\qquad  J_n^{(j)}=w_n^{(j)},\qquad j\geq 2,
\end{eqnarray}
\begin{eqnarray}
\lefteqn{}
&&J_n^{(0)}=(E^2+E+1)^{-1}[v_{n+1}+v_n-p_n(E^2+E+1)^{-1}p_{n+1}],\nonumber\\
&&J_n^{(1)}=q_{n+1}(E^2+E+1)^{-1}(p_{n+1}+p_n),\nonumber\\
&&J_n^{(j)}=w_n^{(j)}(E^2+E+1)^{-1}(p_{n+1}+p_n)+\sum_{l+s=j}w_n^{(l)}w_{n+1}^{(s)}, \qquad j\geq 2
\end{eqnarray}
and 
\begin{eqnarray}
\lefteqn{}
&&J_n^{(0)}=u_n, \qquad J_n^{(1)}=v_nq_{n+1},\nonumber\\
&&J_n^{(j)}=v_nw_n^{(j)}+p_n\sum_{l+s=j}w_n^{(l)}w_{n+1}^{(s)}+\sum_{l+s+m=j}w_n^{(l)}w_{n+1}^{(s)}w_{n+2}^{(m)},
\qquad j\geq 2
\end{eqnarray}
\setcounter{equation}{0}
\section{Conclusions and discussions}
As is well known, discrete zero curvature representation and infinitely many conservation laws are two important integrable properties for a discrete lattice system. Specially, there is less work on the infinitely many conservation laws for the 2+1 dimensional lattice hierarchy.  
In this paper, we describe the discrete zero curvature representations for several 2+1 dimensional lattice hierarchies proposed by Blaszak and Szum. By means of solving the corresponding discrete spectral equations,
we also demonstrate the existence of infinitely many conservation laws for these 2+1 dimensional lattice hierarchies
and derive the corresponding conserved densities and associated fluxes. To our knowledge, the explicit constructions of infinitely many conserved quantities associated with 2+1 dimensional lattice hierarchy are remarkable. It is remarked that equations (2.6) and (2.20) become the so-called differential-difference KP equation by a simple variable transformation [12] and 2+1 dimensional B-S lattice-field equations (2.25)-(2.27) and (2.34)-(2.36) are the generalizations
of 1+1 dimensional Blaszak and Marciniak lattice-field equations [13]. Equation (2.25) is also viewed as a two-dimensional generalization of a lattice in Ref. [14]. Some integrable 
properties on equations (2.6), (2.20) and (2.25), such as lump solution, several Lie symmetries, B\"{a}cklund transformation have been derived in [12]. Here we present their infinitely many conservation laws. So, their integrability is further confirmed. It is also interesting to note that 1+1 dimensional reductions of equations (2.27), (2.34) and (2.35) are new 1+1 dimensional lattice equations.\\  
\noindent
{\bf {Acknowledgments}}\\
This work was supported by Research Grant Council of Hong Kong (HKBU2047/02P).
\end{document}